\input harvmac

 \vskip.4in \centerline{\bf Supersymmetry, the Cosmological Constant, and a}
 \centerline{\bf Theory of Quantum Gravity in Our Universe}
  \vskip.3in

\centerline{\it T. Banks}\vskip.1in \centerline{Department of
Physics and Institute for Particle Physics} \centerline{University
of California, Santa Cruz, CA 95064} \centerline{and}
\centerline{Department of Physics and Astronomy, NHETC}
\centerline{Rutgers University, Piscataway, NJ 08540} \vskip.3in
\centerline{E-mail: \ banks@scipp.ucsc.edu} \vskip.5in

\centerline{\bf ABSTRACT}

There are many theories of quantum gravity, depending on
asymptotic boundary conditions, and the amount of supersymmetry.
The cosmological constant is one of the fundamental parameters
that characterize different theories.  If it is positive,
supersymmetry must be broken. A heuristic calculation shows that
a cosmological constant of the observed size predicts
superpartners in the TeV range. This mechanism for SUSY breaking
also puts important constraints on low energy particle physics
models.  This essay was submitted to the Gravity Research
Foundation Competition and is based on a longer article, which
will be submitted in the near future.

\vfill\eject
 Superstring Theory (ST) is our most successful
attempt at constructing a quantum theory of gravitation.  The
advances of the Duality Revolution\ref\dr{ J.~Schwarz, {\it
Lectures on Superstring and M-theory Dualities Given at ICTP
Spring School and at TASI}, Published in Nucl. Phys. Proc. Suppl.
55B,1,(1997), hep-th/9607201. } gave us detailed mathematical
evidence for the nonperturbative existence and consistency of the
theory. Ironically, they also told us that its name is misleading
because it emphasizes particular asymptotic regions of a
collection of continuous moduli spaces of theories. A better name
would be {\it Supersymmetric Quantum Theories of Gravity
(SQUIGITS)}.

Indeed, the most cogent statement of the results of the Duality
Revolution is that the principles of supersymmetry (SUSY) and
quantum mechanics imply the existence of these moduli spaces of
theories and of certain extended objects in them, whose tension
can be calculated exactly.  One then sees that in certain limiting
regions of moduli space, strings of tension much less than the
Planck scale exist, and one is led to expect a perturbative theory
of strings.  The existing formalism of perturbative superstring
theory is a brilliant confirmation of these general arguments.
Almost all known perturbative string expansions can be derived
from arguments of this sort.  The perturbation expansions allow us
to calculate many quantities whose value does not follow from
SUSY.  More remarkably, in many cases, they can be used to obtain
a completely nonperturbative formulation of the theory.  The
latter examples go under the names of Matrix
Theory\ref\bfss{T.~Banks, W.~Fischler, S.~Shenker, L.~Susskind
{\it M-theory as a Matrix Model: A
Conjecture},Phys.Rev.D55:5112-5128,1997, hep-th/9610043 } and the
AdS/CFT correspondence\ref\oa{ J.~Maldacena, {\it The Large N
Limit of Superconformal Field Theories and Supergravity}, Adv.
Theor. Math. Phys. 2, 231, (1998), Int J. Theor. Phys. 38,
(1999); S.~Gubser, I.~Klebanov, A.~Polyakov, {\it Gauge Theory
Correlators From Non Critical String Theory } ,
Phys.Lett.B428:105-114,1998, hep-th/9802109 ;E.~Witten, {\it
Anti-De Sitter Space and Holography },
Adv.Theor.Math.Phys.2:253-291,1998, hep-th/9802150. }.

Two points on a connected moduli space of such theories can be
considered part of the same system because any physical observable
of one can be recovered with arbitrary accuracy in terms of
measurements done in the other.  But this is no longer true if we
try to compare theories on different moduli spaces\ref\isovac{
T.~Banks, {\it On Isolated Vacua and Background Independence} ,
hep-th/0011255; {\it A Critique of Pure String Theory}, Talk
Given at Strings 2002, Cambridge, UK, July 2002; {\it A Critique
of Pure String Theory: Heterodox Opinions of Diverse Dimensions},
manuscript in preparation. }. We seem to be presented with a
plethora of different consistent theories of quantum gravity, all
of which are exactly supersymmetric and none of which describe
the real world.  It behooves us to search for criteria, which
would help us to understand how to construct a theory of the
world, and to explain why our world is not described by a point
on one of these moduli spaces of consistent theories.

An important general principle that emerges\foot{This principle
could have been declared earlier, on the basis of black hole
physics. However, only the mathematically rigorous formulation of
the SUSic theories, particularly the AdS/CFT correspondence,
gives us confidence that it is correct.} from our rigorous
understanding of supersymmetric theories of quantum gravity is
the principle of Asymptotic Darkness: {\it The high energy
spectrum of a theory of quantum gravity is dominated by black
holes\ref\banksah{ O.~Aharony, T.~Banks, {\it Note On the Quantum
Mechanics of M-theory}, JHEP 9903:016, (1999), hep-th/9812237.
}.  All scattering amplitudes at sufficiently large values of the
kinematic invariants are dominated by black hole
production\ref\penetal{R.~Penrose {\it unpublished} 1974; P.D.~
d'Eath, P.N.~ Payne,{\it Gravitational Radiation in High Speed
Black Hole Collisions, 1,2,3} Phys. Rev. D46, (1992), 658, 675,
694; D.~Amati, M.~Ciafaloni, G.~Veneziano, {\it Classical and
Quantum Gravity Effects from Planckian Energy Superstring
Collisions}, Int. J. Mod. Phys. A3, 1615, (1988), {\it Can
Space-Time Be Probed Below the String Size?}, Phys. Lett. B216,
41, (1989). H-J.~Matschull, {\it Black Hole Creation in
($2+1$)-Dimensions}, Class. Quant. Grav. 16, 1069, (1999);
T.~Banks, W.~Fischler, {\it A Model for High Energy Scattering in
Quantum Gravity}, hep-th/9906038; D.~Eardley, S.~Giddings, {\it
Classical Black Hole Production in High-Energy Collisions}, Phys.
Rev. D66, 044011, 2002, gr-qc/0201034. }.} The famed UV/IR
connection\ref\wittsuss{L~.Susskind, E.~Witten, {\it The
Holographic Bound in Anti de Sitter Space}, hep-th/9805114.}
follows from this principle\foot{as does the even more famous
Holographic Principle.  One can attempt to probe short distances
in order to demonstrate the volume extensive density of states we
expect from quantum field theory (QFT). The production of black
holes prevents us from doing this, and instead presents us with
an area extensive spectrum of states.} : high energy states take
up large regions in space, and have low curvature external
gravitational fields. This connection is the key to understanding
that isolated vacuum states or theories with different values of
the cosmological constant are not connected. The traditional
notion of vacuum state in QFT is an infrared notion.  Two vacua
of the same QFT have identical high energy behavior, but this is
false for states with different values of the cosmological
constant.

For negative cosmological constant, the evidence for this
statement comes from AdS/CFT. In these systems, the value of the
cosmological constant in Planck units is determined by an integer
$N$.  $N$ determines the number of degrees of freedom of the
conformal field theory whose boundary dynamics defines quantum
gravity in the bulk of AdS space.  For $5$ dimensional AdS spaces
the relevant theories are conformally invariant supersymmetric
gauge theories and $N$ is the rank of the gauge group.  Large $N$
corresponds to small cosmological constant, $\Lambda$.  {\it
AdS/CFT shows us that the value of $\Lambda$ is a discrete choice
that we make in defining the theory, rather than a computable
quantity in the low energy effective action.} $\Lambda$
determines the density of {\it high} energy states of the theory.

For positive $\Lambda$ the evidence is less compelling since we do
not yet have a mathematical quantum theory of de Sitter (dS)
space. Fischler\ref\willy{ W.~Fischler, {\it Taking de Sitter
Seriously}. Talk given at {\it The Role of Scaling Laws in
Physics and Biology (Celebrating the 60th Birthday of Geoffrey
West)}, Santa Fe Dec. 2000, and unpublished. } and the
author\ref\tbfolly{T.~Banks, {\it QuantuMechanics and CosMology},
Talk given at the festschrift for L. Susskind, Stanford
University, May 2000; {\it Cosmological Breaking of
Supersymmetry?}, Talk Given at Strings 2000, Ann Arbor, MI, Int.
J. Mod. Phys. A16, 910, (2001), hep-th/0007146} suggested that
the Bekenstein-Gibbons-Hawking entropy of dS space be interpreted
as the logarithm of the number of quantum states in the Hilbert
space defining quantum dS gravity. Evidently, $\Lambda$ is then a
discrete parameter chosen by the theorist, just as it is in the
AdS systems.  The existence of a maximal size black hole with
entropy less than the dS entropy, together with the Bekenstein
bound on the entropy of general localized systems by the entropy
of black holes, then implies that a finite number of states
suffices to describe any conceivable measurement in dS space.

The new role of $\Lambda$ as a fundamental parameter, suggests
that we give up the attempt to explain its value by other than
anthropic means.  Rather, we should attempt to calculate
everything else in the theory, as a function of $\Lambda$, in
Planck units, and use one experiment to determine that pure
number. The opportunity to explain the conundrum of vacuum
selection now presents itself. A unitary quantum theory of dS
space cannot be SUSic.  Thus, the choice of a finite dimensional
Hilbert space for the quantum theory, breaks SUSY. The question is
by how much. Low energy field theory suggests a gravitino mass
that scales like $\Lambda^{1/2}$ in Planck units. I have presented
a quantum calculation, which suggests an enhancement
to\ref\sushor{ T.~Banks, {\it Breaking SUSY at the Horizon},
hep-th/0206117} \eqn\subreak{m_{3/2} \sim \Lambda^{1/4}.}

The key to the calculation is the fact that the $\Lambda
\rightarrow 0$ limit of the dS theory is a SUSic, R-symmetric
theory\tbfolly\ . R symmetry violating terms are induced in the
low energy effective Lagrangian by the dS background. These then
lead to spontaneous violation of SUSY, and a gravitino mass. The
leading contribution to these R violating terms comes from
Feynman diagrams where a single gravitino line propagates out to
the horizon and interacts with the large number of degenerate
quantum states that a static observer sees there. The graph is
suppressed by $e^{- c m_{3/2}R}$ from the gravitino propagator,
where $R = \Lambda^{- {1 \over 2}}$ is the radius of the horizon.
I argued that there was a compensating factor $e^{b \over
m_{3/2}}$ from the interaction with the large set of degenerate
states on the horizon.   Self consistency then leads to the
scaling law $m_{3/2} \sim \Lambda^{1/4}$.  If the two exponential
terms don't cancel exactly, leading to power law corrections, then
the assumption of a small gravitino mass leads to a very large
mass and vice versa.  Of course, we really need a complete
mathematical theory of the dS horizon to construct a reliable
version of this argument.

The phenomenological consequences of the calculation are
significant. The consistent SUSY vacua are banished from the
theory of the world by the assumption that theory has a finite
number of states. The dimension of the dS space is probably fixed
to be $4$ because this is the only dimension in which Superstring
Theory can have a low energy Lagrangian with small deformation
that supports a dS solution. The limiting SUSY theory cannot have
any moduli. The value of superpartner masses that solves the
Standard Model hierarchy problem is predicted in terms of the
cosmological constant. There are also more detailed constraints
on the possible forms of the low energy supersymmetric
Lagrangian\ref\susypheno{T.~Banks, {\it The Phenomenology of
Cosmological SUSY Breaking}, hep-ph/0203066. }.

One problem that remains to be solved is the construction of
machinery for finding all possible isolated Super Poincare
invariant theories of quantum gravity in four dimensions. A
second is the development of a mathematical theory of horizon
dynamics, which will provide a firm foundation for the
calculation of supersymmetric mass splittings in terms the
cosmological constant.

\listrefs
\end